\documentclass[12pt]{article}
\usepackage{latexsym}
\usepackage{mathcomp}
\usepackage{graphicx}
\usepackage{graphics}
\usepackage{epstopdf}
\usepackage{epsfig}
\usepackage{graphicx}
\usepackage{float}
\usepackage{amsmath}
\usepackage{amsthm}
\usepackage{amssymb}
\usepackage{amscd}
\usepackage{amsfonts}
\usepackage{makeidx}
\usepackage{arcs}

 \newcommand{\beqn}{\begin{eqnarray}}
 \newcommand{\eeqn}{\end{eqnarray}}
 \newcommand{\be}{\begin{equation}}
 \newcommand{\ee}{\end{equation}}
 \newcommand{\ba}{\begin{array}}
 \newcommand{\ea}{\end{array}}

 \newcommand{\re}{\ref}

 \newcommand{\bfr}{\begin{flushright}}
 \newcommand{\efr}{\end{flushright}}
 \newcommand{\bfl}{\begin{flushleft}}

 \newcommand{\efl}{\end{flushleft}}

 \newcommand{\ds}{\displaystyle}

 \newcommand{\br}{|\kern-.25em|\kern-.25em|}
 \newcommand{\brr}{{|\kern-.15em|\kern-.15em|\kern-.15em}\,}
 \newcommand{\ddd}{\st{.\kern-.07em.\kern-.07em.}}

\newcommand{\bo}{{\hfill\loota}}
\newcommand{\loota}{\hbox{\enspace{\vrule height 7pt depth 0pt width
      7pt}}}

 \def\N{\mathbb{N}}                             

\def\R{\mathbb{R}}                              
 
 \def\Z{\mathbb{Z}}                                 
 
 \newtheorem{theorem}{Theorem}[section]

 \newtheorem{lemma}[theorem]{Lemma}
 
 \newtheorem{remark}[theorem]{Remark}

\begin{document}
\title{Rayleigh-Bloch waves trapped by a periodic perturbation: exact solutions}
\author{
\large{A.  Merzon$^1$},\\
\large{ P.  Zhevandrov$^2$},\\
\large{M.I. Romero Rodr\'iguez$^3$}\\
\large{and J.E. De la Paz M\'endez$^4$}\\
{\small{\it $^1$ Instituto de F\'\i sica y  Matem\'aticas, Universidad Michoac{a}na}},\\[-2mm]
{\small  Morelia, Michoac\'{a}n, M\'{e}xico}\\[-2mm]
{\small{\it $^2$ Facultad de  Ciencias F\'\i sico-Matem\'aticas, Universidad Michoac{a}na}},\\[-2mm]
{\small  Morelia, Michoac\'{a}n, M\'{e}xico}\\[-2mm]
{\small{\it $^3$ Universidad Militar Nueva Granada}},\\[-2mm]
{\small{Bogot\'a, Colombia}},\\[-2mm]
{\small{\it $^4$ Facultad de Matem\'aticas, Universidad Aut\'onoma de Guerrero}},\\[-2mm]
{\small{Chilpancingo, Guerrero, M\'exico}}\\[-2mm]
{\small{\it E-mails}: anatoli@gmail.com,}
{\small pzhevand@gmail.com,}\\[-2mm] {\small maria.romeror@unimilitar.edu.co,} {\small jeligio12@gmail.com}}
\date{}
\maketitle
\begin{center}
\date{\today}
\end{center}
\begin{abstract}
Exact solutions describing the Rayleigh-Bloch waves for the two-dimensional Helmholtz equation are constructed in the case when the refractive index is
a sum of a constant and a small amplitude function which is periodic in one direction and of finite support in the other. These solutions are quasiperiodic along the structure and exponentially decay in the orthogonal direction. A simple formula for the dispersion relation of these waves is obtained.

\end{abstract}


\section{Introduction}

The appearance of trapped modes in unbounded domains
under perturbations has attracted a
lot of attention in both physical and mathematical literature in the recent past (see, e.g., the books \cite{Hurt, Lond, ExH} and references therein). A classical example is the appearance of a bound state (which is the term for a trapped mode in quantum mechanics) for the one-dimensional Schr\"odinger operator perturbed by a small potential well \cite{Land, Sim76}. In this situation the unperturbed problem possesses a purely continuous spectrum which coincides with the positive ray, and,
under a perturbation by a potential well $\varepsilon V(x)$ with $\varepsilon\to+0$ and a smooth  compactly supported function $V(x)$ such that $\int V(x)\,dx<0$, the cut-off  of the continuous spectrum (that is, the origin), gives rise
to an eigenvalue   at a distance $\sim\varepsilon^2$ to the left of it.
In terms of waves trapped by a one-dimensional perturbation of the Helmholtz equation in dimension 2 (this equation will be our setting in the present paper), this result can be rephrased as follows. Consider the Helmholtz equation
\begin{equation}\label{Helm}
-\Delta\Psi=\frac{\omega^2}{c^2(x,y)}\Psi,~~~\Delta=\partial_{x}^2+\partial_{y}^2,~~x,y\in\R,
\end{equation}
in the plane and assume that $c^{-2}(x,y)=1+\varepsilon f(x),\ \varepsilon\to+0$, where $f(x)$ is such that $\ds\int f(x)\ dx>0$, is smooth, and vanishes for $|x|>R$, and $\omega^2$ is the spectral parameter. One can separate the variable $y$ assuming that
\begin{equation}\label{HEL}
\Psi(x,y)=e^{i\beta y}\ \Psi_{0}(x)
\end{equation}
with $\beta\neq0$ (oblique incidence). Substituting in (\ref{Helm}) and dividing by ${\rm exp} (i\beta y)$ we obtain for $\Psi_0$ the following equation
$$
-\Psi''_{0}(x)+\beta^2\Psi_{0}(x)=\omega^2(1+\varepsilon f(x))\Psi_{0}(x).
$$
The continuous spectrum of this problem is the ray $\omega^2\geq \beta^2$ and the cut-off $\omega^2=\beta^2$ gives rise, under the perturbation, to an eigenvalue $\omega^2=\beta^2-\mu^2$ with $\mu$ analytic in $\varepsilon$
and given by
\begin{equation}\label{MU}
\mu=\frac{\varepsilon\beta^2}{2}\int f(x)\ dx+O(\varepsilon^2).
\end{equation}

Here we will be interested in a generalization of this result to the case of a weak periodic perturbation
\begin{equation}\label{cVAR}
c^{-2}(x,y)=1+\varepsilon f(x,y),
\end{equation}
where $f(x,y)$ is smooth, $T$-periodic with respect to $y$, $f(x,y+T)=f(x,y)$, and vanishes for $|x|>R$.

It is well-known that a periodic structure can support the so-called {\em Rayleigh-Bloch} (RB) {\em waves} which are solutions of the Helmholtz equation quasiperiodic in $y$ and decaying as $|x|\to \infty$. These waves were studied numerically as well as analytically for the settings of a half-plane with a periodic boundary (see e.g. \cite{W1, Bon, Naz} and references therein) and in the setting of a periodic array of solid obstacles (see e.g. \cite{Por Ev, Pet Mey, McI} and references therein). Therefore it is natural to investigate whether the mechanism of their appearance persist for a weak periodic perturbation of the refractive index (\ref{cVAR}). The construction of RB waves reduces to the solution of a boundary value problem in the strip
$$
\Pi=\R\times(0,T)
$$
for the Helmholtz equation with quasiperiodic boundary conditions; that is, we come to a waveguide problem.

Problems of trapped modes in waveguides were quite extensively studied in numerous publications. We note the papers  \cite{Exner95, Bulla97}, where weakly perturbed quantum waveguides (i.e., perturbed Laplace operator with Dirichlet boundary conditions) were studied by means of the Birman-Schwinger technique (see \cite{ExH}).

Our goal in the present paper is to construct explicit solutions describing  the RB waves for the wave speed given by (\ref{cVAR}). They have the form
\begin{equation}\label{PBY}
\Psi(x,y)=e^{i\beta y}\psi(x,y),
\end{equation}
where  $\psi$ is $T$-periodic in $y$ and decays as $|x|\to\infty$, and one can assume that $-\pi/T<\beta\leq \pi/T$ by the periodicity of $\psi$. The principal difficulty of this problem (as well as the others mentioned above) consists in the fact that the nonperturbed problem does not possess trapped modes, so that the standard regular perturbation theory is not applicable. We show that solutions (\ref{PBY}) exist for certain values of $\omega^2$ below the cut-off and provide exact explicit formula for the dispersion relation (i.e. the dependence of $\omega$ on $\beta$) and RB waves themselves in the form of infinite convergent series of powers of $\varepsilon$.

Our approach uses the main idea of the of the Birman-Schwinger method (i.e., the reduction of differential equations to integral ones) with simplifications and modifications (cf. \cite{ACZ}), \cite{MIR Zhev}). Our main result is the asymptotics of the frequencies of RB waves,
\begin{equation}\label{mf}
\omega^2=\beta^2-\mu^2,~~~~\mu=\frac{\varepsilon\beta^2}{2T}\int\int_{\Pi} f(x,y)\ dx dy+O(\varepsilon^2),\qquad 0<|\beta|<\pi/T.
\end{equation}
This result appears to be new and passes into (\ref{MU}) when $f$ does not depend on $y$. When the cut-off $\beta^2$ approaches zero, the frequency of the RB wave also approaches zero and the RB wave becomes a normally incident wave not decaying as $|x|\to\infty$; thus $\beta=0$ is excluded in (\ref{mf}) just as in the case of a one-dimensional perturbation one has to consider oblique waves in order to obtain the trapping phenomenon. We also exclude the case $\beta=\pi/T$ since the thresholds of the continuous spectrum (see formula (\ref{S1}) below) coalesce for $n=0$ and $n=-1$; this case requires a separate treatment.

\section{Formulation}
\setcounter{equation}{0}

Consider the Helmholtz equation
\begin{equation}\label{h1}
-\Delta\Psi(x,y)=\ds\frac{\omega^2}{c^2(x,y)}\Psi(x,y),~(x,y)\in\R^{2},~\omega\in\R,
\end{equation}
with a real smooth $T$-periodic in $y$ refractive index,
\begin{equation}\label{h2}
c(x,y+T)=c(x,y),~(x,y)\in\R^{2}
\end{equation}
(see Fig.\;1).

\vspace{1cm}
\begin{center}
\includegraphics[scale=0.3]{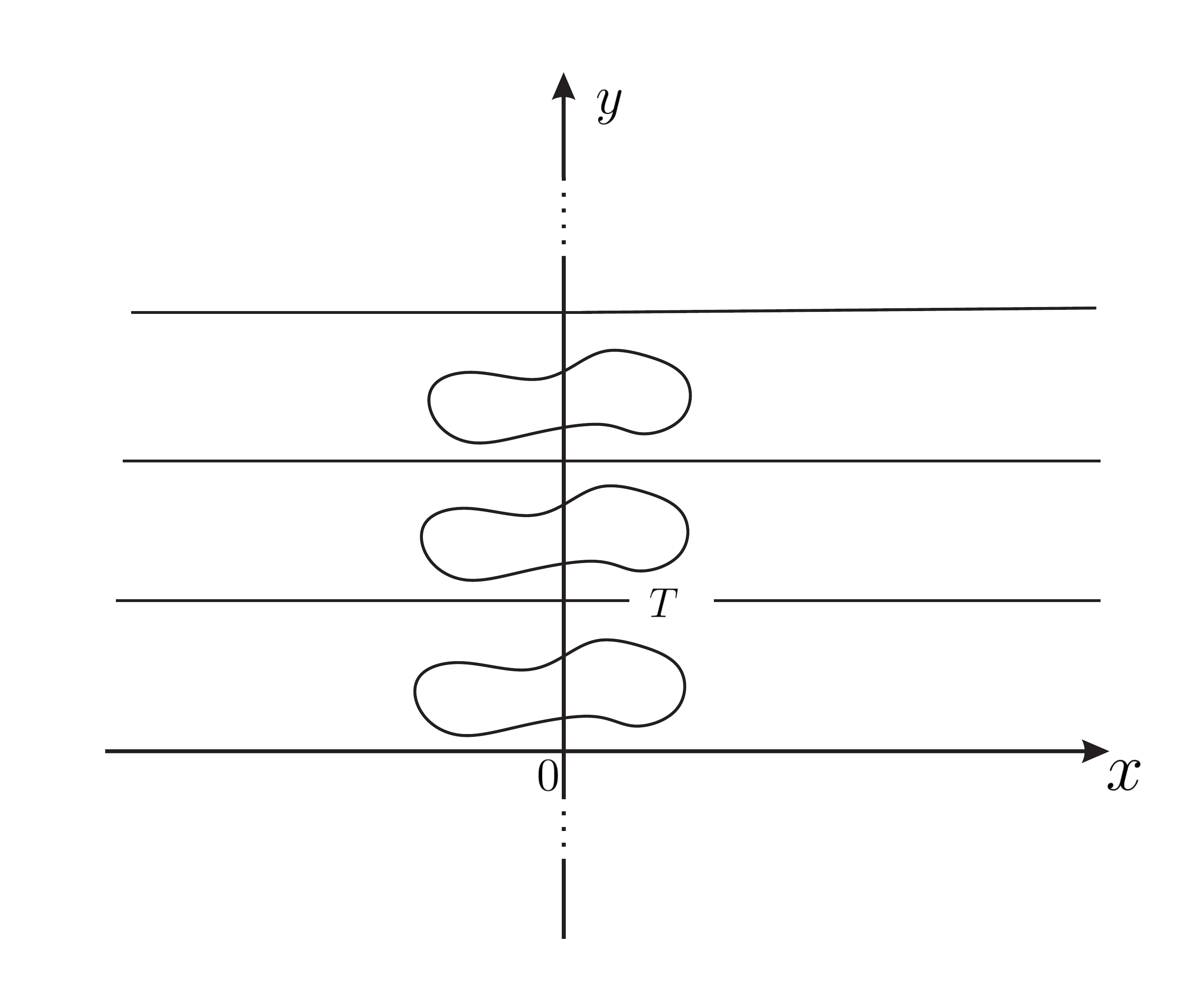}
\end{center}
\centerline{Figure 1}

\vspace{.5cm}
Solutions of this equation that are quasiperiodic in $y$,
\begin{equation}\label{CP1}
\Psi(x,y+T)= e^{i\beta T}\Psi(x,y),\qquad 0<|\beta|<\pi/T,
\end{equation}
bounded in the whole plane and decaying as $|x|\to\infty$ are called {\it Rayleigh-Bloch waves} (RB waves) and exist only for certain values of the spectral parameter $\omega$.

We will construct RB waves in the case of a weakly perturbed refractive index:
\begin{equation}\label{C1}
c^{-2}(x,y)= 1+\varepsilon f(x,y),~~0<\varepsilon\ll 1,
\end{equation}
where $f$ is a smooth $T$-periodic in $y$ function vanishing for $|x|>R$ for some $R>0$.

Obviously, in order to construct the RB waves it is sufficient to find solutions of the following spectral BVP in the strip
$\Pi:=\R\times(0,T)$:
\begin{equation}\label{hh2}
-\Delta\Psi(x,y)=\omega^{2}\Big(1+\varepsilon f(x,y)\Big)\Psi(x,y),\qquad (x,y)\in\Pi,
\end{equation}
\begin{equation}\label{bc1}
\Psi(x,T)=e^{i\beta T}\Psi(x,0),~~~\Psi_{y}(x,T)=e^{i\beta T}\Psi_{y}(x,0)
\end{equation}
\begin{equation}\label{bc3}
\Psi(x,y)\to 0,~~~|x|\to\infty,~~y\in\R.
\end{equation}
We can fix $T=2\pi$ without loss of generality, and we will assume that
\begin{equation}\label{CP}
\int\int_{\Pi} f\,dxdy>0.
\end{equation}
Note that the continuous spectrum of problem (\ref{hh2})-(\ref{CP}) is the ray $\omega^2\geq\beta^2$ and coincides with the continuous spectrum of the unperturbed problem ($\varepsilon=0$). The latter
 is divided by the cut-offs $\omega^2=(n\pm|\beta|)^2,~n=1,2,3,\dots,$ into intervals where its multiplicity is constant;
 this multiplicity is equal to 2 on the first interval $\beta^2<\omega^2<(1-|\beta|)^2$ and is augmented by 2 when passing through the next threshold.
This follows from the explicit form of the plane waves $\exp\{i(\beta+n)y+ikx\}$ satisfying (\ref{hh2})-(\ref{CP}) for $\varepsilon=0$
with $\omega$ given by
\begin{equation}\label{S1}
\omega^2=k^2+\big(\beta+ n\big)^{2}, ~n\in\Z, ~k\in\R.
\end{equation}

\section{Trapped waves}
\setcounter{equation}{0}

We seek the solution of the  spectral problem
\begin{equation}\label{PK}
\left.
\begin{array}{c}
-\Delta\Psi(x,y)=\omega^2\Big(1+\varepsilon f(x,y)\Big)\Psi(x,y),\quad (x,y)\in\Pi,
\\
\Psi(x,2\pi)=e^{2\pi i\beta }\Psi(x,0),\quad
\Psi_{y}(x,2\pi)=e^{2\pi i\beta}\Psi_{y}(x,0),
\\
\Psi(x,0)\to 0,\quad |x|\to\infty,\quad y\in\R,
\end{array}
\right.
\end{equation}
for $\beta$ such that $0<|\beta|<1/2$.
We  seek the eigenvalue $\omega^2$ in the form
\begin{equation}\label{01}
\omega^2=\beta^2-\mu^2,
\end{equation}
where
\begin{equation}\label{mu}
\mu>0,~~\mu\to0,~{\rm as}~~ \varepsilon\to0.
\end{equation}
\begin{remark}\label{rm}
The function $\Psi$ in (\ref{PK}) depends on three argument $x, y, \mu$. In the following we will sometimes omit the dependence of $\Psi$ on $\mu$.

\end{remark}

In (\ref{PK}) the function $\Psi(x,y)$ belongs to the space $H_{\beta}^{1} (\Pi)$ which is the completion of the space of smooth functions satisfying the boundary conditions from (\ref{PK}) and vanishing for large $|x|$ with to respect to norm
\begin{equation}\label{n1}
\parallel\Psi\parallel_{H^{1}_{\beta}(\Pi)}=\parallel\Psi\parallel_{L^{2}(\Pi)}+\parallel\nabla\Psi\parallel_{L^{2}(\Pi)}.
\end{equation}
We understand  problem (\ref{PK}) in the sense of the integral identity: for all $\Phi\in H_{\beta}^{1} (\Pi)$
\begin{equation}\label{it1}
\int_{\Pi} \nabla\Psi\cdot\nabla\overline{\Phi}\ dx dy=\omega^2\int_{\Pi}c^2(x,y)\ \Psi\overline{\Phi}\ dxdy.
\end{equation}
This integral identity easily follows from (\ref{PK}) after multiplying by $\overline{\Phi}$ and integrating over $\Pi$:
\begin{eqnarray}\label{ii1}
\begin{array}{ll}
\ds\int_{\Pi}\Delta\Psi\overline{\Phi}\ dxdy=\int_{\Pi} \nabla\cdot (\overline{\Phi}\ \nabla\Psi)\ dxdy\ -\int_{\Pi}\nabla \Psi\cdot\nabla\overline{\Phi}\ dxdy\\\\\hspace{2.6truecm}=\ds\int_{y=2\pi} \Psi_{y}\ \overline{\Phi}\ dx\ -\int_{y=0} \Psi_{y}\ \overline{\Phi}\ dx\ -\int_{\Pi} \nabla\Psi\cdot\nabla\overline{\Phi}\ dxdy.
\end{array}
\end{eqnarray}
The first two terms in the last expression cancel out due to the boundary conditions in (\ref{PK}). Moreover, the same argument shows that the operator corresponding to the problem (\ref{PK}) is self-adjoint.
\begin{remark}\label{sa}
Strictly speaking, this calculation shows that the operator generated by the sesquilinear form in the left-hand side of (\ref{it1})  is self-adjoint in $H_{1}^{\beta}(\Pi)$ (see \cite{ExH}). Nevertheless the indication of this concrete space is not very important in what follows since we will construct a classical smooth finite energy solution of (\ref{PK}) (see (\ref{2B})).
\end{remark}
For $\beta$ satisfying $0<|\beta|<1/2$, we seek $\mu(\beta)$  and a nontrivial $\Psi$ satisfying (\ref{PK}) such that $\Psi$ decays as $|x|\to\infty$.
We seek $\Psi$ in the following form:
\begin{equation}\label{PS1}
\Psi(x,y,\mu)=e^{i\beta y}\sum_{n\in\Z} \Psi_{n}(x,\mu)\ e^{ i ny},~~(x,y)\in\Pi.
\end{equation}

Substituting (\ref{PS1}) in (\ref{PK}), we obtain
\begin{equation}\label{PS5'}
\Big(\beta^2-\mu^2\Big)\Big(1+\varepsilon f(x,y)\Big)\ds\sum_{n\in\Z} \Psi_{n}(x)\ e^{i\beta y+ i ny}=-\ds\sum_{n\in\Z} \Bigg(\Psi''_{n}(x)-\Big(\beta+  n\Big)^{2}\ \Psi_{n}(x)\Bigg)\ e^{i\beta y+ i ny}.
\end{equation}
Obviously, the boundary conditions in (\ref{PK}) are satisfied automatically for the function (\ref{PS1}) (of course we assume that the series converge sufficiently rapidly, which we will prove below).

\medskip

Multiply  both parts of (\ref{PS5'}) by $\exp\{-i(\beta+m) y\}$. We obtain
\begin{equation}\label{i1}
\Big(\beta^2-\mu^2\Big)\Big(1+\varepsilon f(x,y)\Big)\ds\sum_{n\in\Z} \Psi_{n}(x)\ e^{ i(n-m)y}=-\ds\sum_{n\in\Z} \Bigg(\Psi''_{n}(x)-\Big(\beta+  n\Big)^{2}\ \Psi_{n}(x)\Bigg)\ e^{ i(n-m) y}.
\end{equation}
Integrating   (\ref{i1}) over $y$ from $0$ to $2\pi$, we come to
\begin{equation}\label{i4}
-\Psi''_{m}(x)+\Big(\mu^2+2\beta m+ m^2\Big)\Psi_{m}(x)=\varepsilon\Big(\ds\frac{\beta^2-\mu^2}{2\pi}\Big)\ds\sum_{n\in\Z}\Psi_{n}(x)\ f_{m-n}(x), m\in\Z, x\in\R,
\end{equation}
where
\begin{equation}\label{i5}
f_{j}(x)=\int_{0}^{2\pi} f(x,y)\ e^{- i jy}\ dy,~~j\in\Z.
\end{equation}
We seek the solution of  system (\ref{i4}) in the form
\begin{equation}\label{2.15'}
\Psi_{m}= G_{m}\ast A_{m},~~m\in\Z,
\end{equation}
where $A_{m}$ are new unknown functions,
\begin{equation}\label{a01}
\langle A_0\rangle:=\int A_0(x)\,dx=1,
\end{equation}
the asterisk denotes the convolution and, for $m\in\Z$, the functions $G_{m}(x,\mu)$ are the Green functions satisfying
\begin{equation}\label{G}
-G''_{m}(x,\mu)+\Big(\mu^2+2\beta m+ m^2\Big) G_{m}=\delta(x).
\end{equation}
\begin{remark}\label{ra01}
The  condition (\ref{a01}) does not affect the generality of our considerations since an eigenfunction is defined up to a multiplicative constant.
\end{remark}
We have
\begin{equation}\label{gm}
G_{m}(x,\mu)=\frac{1}{2k_{m}}\ e^{-k_{m}|x|},~~x\in\R,
\end{equation}
where
\begin{equation}\label{km}
k_{m}=\sqrt{\mu^2+2\beta m+ m^2}.
\end{equation}
\begin{remark}\label{ram}
Note that for $m\neq0$, $G_{m}(x,\mu)$ is analytic in $\mu$ for
\begin{equation}\label{muo}
|\mu|<\mu_{0}=\sqrt{1-2\beta}
\end{equation}
by (\ref{km}).
\end{remark}
Thus, by (\ref{i4}) we come to the following system for the functions $A_{m}$:
\begin{equation}\label{am}
A_{m}(x)=\varepsilon\Big(\frac{\beta^2-\mu^2}{2\pi}\Big)\sum_{n\in\Z}f_{m-n}(x)\Big(G_{n}(x)\ast A_{n}(x)\Big),~~m\in\Z.
\end{equation}
Rewrite (\ref{am}) as
\begin{equation}\label{amm}
A_{m}(x)=\varepsilon\Big(\frac{\beta^2-\mu^2}{2\pi}\Big)\Bigg[f_{m}(x)\Big(G_{0}(x)\ast A_{0}(x)\Big)+
\sum_{n\neq 0}f_{m-n}(x)\Big(G_{n}(x)\ast A_{n}(x)\Big)\Bigg].
\end{equation}
Represent $G_{0}(x)$ in the form
\begin{equation}\label{go}
G_{0}(x)=G_{r}(x)+\frac{1}{2\mu},~~x\in\R,
\end{equation}
where
\begin{equation}\label{g2}
G_{r}(x)=G_{r}(x,\mu)=\frac{1}{2\mu}\Big(e^{-\mu|x|}-1\Big),~~x\in\R.
\end{equation}
Note that $G_{r}(\cdot,\mu)$ admits an analytic continuation to a neighborhood of the origin (e.g., the one defined by (\ref{muo})) from the ray $\lbrace \mu>0\rbrace$ . Thus,
\begin{equation}\label{g1}
G_0\ast A_0=\frac{1}{2\mu}+G_r\ast A_0.
\end{equation}
Hence  the vector
\begin{equation}\label{g3}
{\bf A}:=(\cdots, A_{-1}, A_{0},A_1,\cdots)
\end{equation}
satisfies the following equation:
\begin{equation}\label{ml}
\big(1-\varepsilon\hat{T}_{\mu}\big){\bf A}=\frac{\varepsilon}{2\mu}\ \frac{\beta^2-\mu^2}{2\pi}\ {\bf f},
\end{equation}
where
$$
{\bf f}:=(\cdots,f_{-1}, f_{0}, f_{1}, \cdots)
$$
and
\begin{equation}\label{2A}
\big(\hat{T}_{\mu}{\bf A}\big)_{m}(x,\mu)=\frac{\beta^2-\mu^2}{2\pi}\Bigg[f_{m}(x)\Big(G_{r}(x,\mu)\ast A_{0}(x)\Big)+\sum_{n\neq 0}f_{m-n}(x)\Big(G_{n}(x,\mu)\ast A_{n}(x)\Big)\Bigg],
\end{equation}
$\big(\hat{T}_{\mu}{\bf A}\big)_{m}$ means the {\it m}th component of the vector $\hat{T}_{\mu}{\bf A}$. Note that $\big(\hat{T}_{\mu}{\bf A}\big)_{m}(\cdot,\mu)$ is analytic in $\mu,~ |\mu|<\mu_0$ by Remark \ref{ram}.

Consider the space $\mathcal{A}$ of vectors
$$
{\bf A}=\big(\cdots, A_{-1}(x), A_{0}(x), A_{1}(x),\cdots\big),
$$
where
\begin{equation}\label{gg1}
A_{j}(x)\in C[-R,R],
\end{equation}
with the norm
$$
\parallel{\bf A}\parallel^{2}=\sum_{j=-\infty}^{\infty}\Big(\sup_{x\in[-R,R]}|A_{j}(x)|\Big)^{2}.
$$
Obviously, $\mathcal{A}$ is a Banach space.

\begin{lemma}\label{OP}
{\bf i)}
The operator $\hat{T}_{\mu}:{\bf A}\to{\bf A}$ given by (\ref{2A}) is bounded uniformly in $\mu$ for $|\mu|\leq\mu_0/2$, where $\mu_0$ is given in (\ref{muo}):
$$
\parallel\hat{T}_{\mu} {\bf A}\parallel^{2}=\sum\Big(\sup_{x\in[-R,R]}|(\hat{T}_{\mu} {\bf A})_{m}{(x)}|\Big)^{2}\leq C\ \sum\Big(\sup_{x\in[-R,R]}|A_m(x)|\Big)^{2}.
$$
{\bf ii)} If $A_{n}(x)$  decay rapidly as $n\to\infty$, i.e.,
$$
\sup_{x\in[-R,R]}|A_{n}(x)|=O\Big(|n|^{-N}\Big),~N\in\Z,
$$
then the components of $\hat{T}_{\mu}{\bf A}$ also  decay rapidly as $n\to\infty$ uniformly in $\mu,~|\mu|\leq\mu_0/2$.

\end{lemma}

{\bf Proof.} {\bf i)} Rewrite the operator $\hat{T}_{\mu}$ in the form
\begin{equation}\label{t2}
(\hat{T}_{\mu} {\bf A})_{m}(x,\mu)=\sum_{n} f_{m-n}(x)\Big(\mathcal{H}_{n}(x,\mu)\ast A_{n}(x)\Big)(x,\mu),~ x\in\R,~ |\mu|\leq\mu_0/2,
\end{equation}
where
\begin{equation}\label{t3}
\mathcal{H}_{0}(x,\mu)=\frac{\beta^2-\mu^2}{2\pi}\ G_{r}(x,\mu),~~~~\mathcal{H}_{n}(x,\mu)=\frac{\beta^2-\mu^2}{2\pi}\ G_{n}(x,\mu),~~n\neq0.
\end{equation}
Note that by (\ref{g2}) and (\ref{gm}) there exists $C>0$ such that
\begin{equation}\label{3.31'}
|\mathcal{H}_{n}(x,\mu)|\leq C,~x\in \R,~ n\in\Z,~|\mu|\leq\mu_0/2.
\end{equation}
Let us bound $(\hat{T}_{\mu} {\bf A})_{m}$. We use the Schur lemma for the discrete convolution.
Denote
$$
\hat{T}_{\mu}{\bf A}={\bf B}=(\cdots,B_{-1},B_{0},B_{1},\cdots).
$$
We have, by the Cauchy-Schwarz inequality,
$$
|B_{m}(x,\mu)|^2=\Big|\sum_{n} f_{m-n}(x)(\mathcal{H}_{n}\ast A_{n})(x,\mu)\Big|^{2}\leq \sum_{n}|f_{m-n}(x)| \sum_{n}|f_{m-n}(x)||(\mathcal{H}_{n}\ast A_{n})(x,\mu)|^2.
$$
Hence
$$
\sup_{x\in[-R,R]}|B_{m}(x,\mu)|^2\leq\Big(\sum_{n}\sup_{x\in[-R,R]}|f_{m-n}(x)|\Big)\sum_{n}\sup_{x\in[-R,R]}|f_{m-n}(x)|\sup_{x\in[-R,R]}|(\mathcal{H}_{n}\ast A_{n})(x,\mu)|^{2}.
$$
Note that by (\ref{go}), (\ref{g2}), (\ref{3.31'})
$$
\sup_{x\in[-R,R]}\Big|\Big(\mathcal{H}_{n}\ast A_{n}\Big)(x,\mu)\Big|^{2}\leq C\Big(\sup_{x\in[-R,R]} |A_{n}| \Big)^{2},~|\mu|\leq\mu_0/2,
$$
since $\mathcal{H}_{n}$ are bounded on $[-R,R]$ uniformly in $n$ and $|\mu|\leq\mu_0/2$. Hence,
$$
\sup_{x\in[-R,R]}|B_{m}(x,\mu)|^2\leq C \Big(\sum_{n}\sup_{x\in[-R,R]}|f_{m-n}(x)|\Big)\sum_{n}\Big(\sup_{x\in[-R,R]}|f_{m-n}(x)| \sup_{x\in[-R,R]}|A_{n}(x)|^2\Big).
$$
Summing the last inequality over $m$ we have
$$
\parallel{\bf B}\parallel_{\mathcal{A}}^{2}\leq C\Big(\sum_{n}\sup_{x\in[-R,R]}|f_{n}(x)|\Big)^{2}\parallel{\bf A}\parallel_{\mathcal{A}}^{2}.
$$
The sum in the last inequality is bounded  because $f_{n}$ are the Fourier coefficients of a smooth function. Statement {\bf i)} is proven.

Statement {\bf ii)} follows from the definition of the operator $\hat{T}_{\mu}$ (\ref{t2}) and the fact that $\mathcal{H}_{n}$ are uniformly
bounded, see (\ref{3.31'}).~~~ $\bo$

\medskip

From this lemma, the solution of equation (\re{ml}) is given by
\begin{equation}\label{2B}
{\bf A}=\frac{\varepsilon}{2\mu}\ \frac{\beta^2-\mu^2}{2\pi}(1-\varepsilon\hat{T}_{\mu})^{-1}\ {\bf f}
\end{equation}
for sufficiently small $\varepsilon>0$.
In particular,
\begin{equation}\label{2C}
 A_{0}=\frac{\varepsilon}{2\mu}\ \frac{\beta^2-\mu^2}{2\pi}\Bigg(\Big(1-\varepsilon\hat{T}_{\mu}\Big)^{-1}{\bf f}\Bigg)_{0}.
\end{equation}
Integrating the last equality over $x$, using (\ref{a01}), and multiplying by $\mu$, we come to an equation for $\mu$:
\begin{equation}\label{2D}
\mu=\frac{\varepsilon}{2}\ \frac{\beta^2-\mu^2}{2\pi}\Big\langle\Big((1-\varepsilon\hat{T}_{\mu})^{-1}\ {\bf f}\Big)_{0} \Big\rangle.
\end{equation}
In the leading term, we have
\begin{equation}\label{2F}
\mu= \frac{\varepsilon\ \beta^2}{4\pi}\int\int_{\Pi} f(x,y)\ dx dy+O(\varepsilon^2).
\end{equation}
Therefore we have found the eigenvalue $\mu$ for  problem (\ref{PK}) and the corresponding eigenfunctions given by (\ref{2.15'}), (\ref{2B}), and (\ref{PS1}).

The fact that $\mu$ is real and positive follows from condition (\ref{CP}), formula (\ref{2F}) and self-adjointness of problem (\ref{PK}) (see Remark \ref{sa}). Thus we have proven the following theorem:
\begin{theorem}\label{B0T}
Let $f$ from (\ref{C1}) be a smooth $2\pi$-periodic in $y$ function, vanishing for $|x|>R$ and satisfying (\ref{CP}).
Then for sufficiently small $\varepsilon>0$ there exists an eigenvalue $\omega^2$ of problem (\ref{PK}) such that
\begin{equation}\label{OM}
\omega^2=\beta^2-\mu^2
\end{equation}
with $\mu$ being the solution of (\ref{2D}) with the leading term given by (\ref{2F}). The corresponding eigenfunction is given  by (\ref{PS1}), (\ref{2.15'}), (\ref{gm}), (\ref{2B}), and  belongs to $H_{\beta}^{1}(\Pi)$ for any fixed real $\beta$, $0<|\beta|<1/2$.
\end{theorem}
{\bf Proof.} First, we clarify the meaning of the expression ``sufficiently small'' in the formulation: $\varepsilon$ should be small enough to guarantee that (\ref{OM}) is positive, i.e., $|\mu|<|\beta|$, and to guarantee that (\ref{muo}) holds, i.e., $|\mu|<\mu_0$. Now the existence of the eigenvalue $\omega$ follows from formula (\ref{2D}) by the Implicit Function Theorem and the analyticity of the right-hand side of (\ref{2D}) in $\varepsilon$ and $\mu$ (see Lemma \ref{OP} {\bf ii)}). The fact that the function $\Psi$ belongs to  $H_{\beta}^{1}$ follows immediately from the estimates
$$
|f_{m}(x)|\leq C_{N}(1+|m|)^{-N},~ \forall N\in\N,
$$
which are valid since  $f_{m}$ are the Fourier coefficients of a smooth function. ~~~$\bo$

\begin{remark}
In fact, our considerations also show the uniqueness of the constructed eigenvalue because any solution of problem (\ref{PK}) from $H_{1}^{\beta}$ can be represented in the form (\ref{PS1}) with an appropriate function $\Psi_{n}$.

\end{remark}

\section{Conclusion}

We have constructed an explicit formula for the RB wave for the two-dimensional Helmholtz equation trapped by a weak periodic perturbation of the wave speed.
This formula has the form of the infinite converges series in powers of the small parameter characterizing the perturbation. In particular, we have obtained a very simple formula (\ref{mf}) for the leading term of the frequency of this wave.


\begin{thebibliography}{}
\bibitem{Hurt} Norman E. Hurt, ``\textit{Mathematical physics of quantum wires and devices}'', Kluwer (2000).

\bibitem{Lond} J. T. Londergan, J. P. Carini, D. P. Murdock, ``\textit{Binding and scattering in two-dimensional systems: applications to quantum wires, waveguides and photonic crystals}'', Springer (1999).

\bibitem{ExH} P. Exner, H. Kova\v{r}\'\i k, ``\textit{Quantum waveguides}'', Springer (2015)

\bibitem{Land} L.D. Landau, E.M. Lifshitz, ``\textit{Quantum mechanics. Non-relativistic theory}'', Pergamon (1958) (\S 45)



\bibitem{Sim76} B. Simon. The bound state of weakly coupled Schr\"odinger operators in one and two dimensions, Ann. Phys., v. {97}, pp. 279-288 (1976).


\bibitem{W1} C.H. Wilcox. ``\textit{Scattering theory for diffraction gratings}'', Springer (1984)

\bibitem{Bon} A.S. Bonnet-Bendhia, F. Starling. Guided waves by electromagnetic gratings and non-uniqueness examples for the diffraction problems. Math. Meth. Appl. Sci. v. 17, pp. 305-338 (1994)

\bibitem{Naz} I.V. Kamotski, S.A. Nazarov. Trapped modes, surface waves and wood's anomalies for gently sloped periodic boundaries. CR Acad. Sci. Paris, v. 328, Ser. II b, pp. 423-428 (2000)

\bibitem{Por Ev} R. Porter and D. V. Evans. Rayleigh-Bloch surface waves along periodic gratings and their connection with trapped modes in waveguides,  J. Fluid Mech., v. 386, pp. 233-258 (1999).

\bibitem{Pet Mey} M. A. Peter and M. H. Meylan. Water-wave scattering by a semi-infinite periodic array of arbitrary bodies. J. Fluid Mech.,  v. 575, pp. 473-494 (2007).

\bibitem{McI} P. McIver, C. M. Linton, and M. McIver. Construction of trapped modes for wave guides and diffraction gratings. Proc. Roy. Soc. A , v. 454, pp. 2593-2616 (1998).


\bibitem{Exner95} P. Duclos and P. Exner. Curvature-induced bound states in quantum waveguides in two and three dimensions. Rev. Math. Phys., v. 7, pp. 73-102 (1995).

\bibitem{Bulla97} W. Bulla, F. Gesztesy, W. Renger, and B. Simon. Weakly coupled bound states in quantum waveguides,  Proc. Am. Math. Soc., v. 125, pp. 1487-1495 (1997)



\bibitem{ACZ} H. Aya, R. Cano, P. Zhevandrov. Scattering and embedded trapped modes for an infinite non-homogeneous. Timoshenko beem. J. Engng. Maths, v. 77, pp. 87-104 (2012)

\bibitem{MIR Zhev} M.I. Romero Rodr\'\i guez, P. Zhevandrov. Trapped modes and scattering for oblique waves in a two-layer fluid. J. Fluid Mech., vol. 753, pp. 427-447 (2014)







\end{thebibliography}
\end{document}